\documentstyle[aps,prb,epsfig,twocolumn]{revtex}

\begin{document}
\title{Electron-Hole Correlations and Optical Excitonic Gaps in 
       Quantum-Dot Quantum Wells: Tight-Binding Approach}
\author{Rui-Hua Xie and  Garnett W. Bryant
       \footnote{Corresponding author: \\ 
                 Phone: (301) 975-2595; 
                   Fax: (301) 990-1350; 
                E-mail: garnett.bryant@nist.gov}}
\address{National Institute of Standards and Technology, 
         Gaithersburg, Maryland 20899-8423}
\author{Seungwon Lee}
\address{Department of Physics, Ohio State University, Columbus, 
         Ohio 43210-1106}
\author{W. Jask\'{o}lski}
\address{Instytut Fizyki, UMK, Grudzi{\c a}dzka 5,  87-100 Tor\'{u}n, Poland}
\date{\today} 
\maketitle

\begin{abstract}
Electron-hole correlation in quantum-dot quantum wells (QDQW's) 
is investigated by  incorporating  Coulomb and exchange interactions 
into an empirical tight-binding model. Sufficient electron and hole 
single-particle states close to the band edge are included in the 
configuration to achieve convergence of the first spin-singlet and 
triplet excitonic energies within a few meV. Coulomb shifts of about 
100 meV and exchange splittings of about  1 meV are found for 
CdS/HgS/CdS QDQW's (4.7 nm CdS core diameter, 0.3 nm HgS well width  
and  0.3 nm to 1.5 nm CdS clad thickness) which have been characterized 
experimentally by Weller and co-workers [ D. Schooss, A. Mews, 
A. Eychm\"{u}ller, H. Weller, Phys. Rev. B, 49, 17072 (1994)]. 
The optical excitonic gaps calculated for those QDQW's are in good 
agreement with the experiment. 

\noindent 
 {\bf PACS (numbers)}: 71.24+q; 71.15.Fv; 71.35.Cc; 73.61.Tm


\end{abstract}

\vspace{1cm}


Semiconductor nanostructures,  from quantum wells, quantum wires
to quantum dots\cite{ph00,kbar01},  have been extensively investigated
due to their remarkable  applications, for example, as fast and 
low noise electronic devices and tunable optoelectronic elements.  
Recently, a class of new and promising hetero-quantum dots, termed 
{\sl quantum-dot quantum wells} (QDQW's),  have been successfully 
synthesized in water, for example, CdS/HgS/CdS\cite{ds94,am94,am96}, 
CdTe/HgTe/CdTe\cite{svk99} and  ZnS/CdS/ZnS\cite{rbl2001}. 
These QDQW's  have internal nanoheterostructures with a quantum-well 
region contained inside the  quantum dot. Meanwhile,  high-resolution 
transmission electron microscopy images\cite{am96,svk99} have shown that 
CdS/HgS/CdS and CdTe/HgTe/CdTe QDQWs are not spherical,  but are 
preferentially truncated tetrahedral particles. However, spherical 
shell particles are commonly  considered to explain experimental  
results\cite{ds94,am94,jwh93,gwb95,wj98,gwb01,bryant02}. 

Numerical calculations\cite{ds94,am94,jwh93,gwb95} on QDQW's have been 
based  generally  on the one-band effective-mass approximation as well 
as the parabolic approximation  for the conduction and valence bands. 
These theoretical studies have demonstrated the remarkable effect of the 
internal well on single-particle electron  and hole energies and pair 
overlaps\cite{ds94,am94,jwh93} and  determined the  ground-state energy 
of an uncorrelated electron-hole pair\cite{ds94,am94}. Moreover, Bryant
\cite{gwb95} determined the contribution of pair correlation  to 
the electronic structure of QDQW's nanosystems. Very recently, 
Jask\'{o}lski and Bryant\cite{wj98} have developed a multi-band theory 
to determine electron, hole and exciton states for QDQW's. Actually, an 
atomic model is essential for QDQW's since the internal wells are no more 
than a few monolayers thick\cite{gwb01,bryant02}. Hence, in this paper,  we 
incorporate for the first time Coulomb and exchange interactions into 
an empirical  tight-binding model to describe QDQW nanocrystals. In 
detail, we shall investigate electron-hole interactions and optical 
excitonic gaps of QDQW's, and report our numerical results by considering  
three CdS/HgS/CdS QDQW's which were characterized experimentally by Weller 
and coworkers\cite{ds94}. Comparison with the experiment shows that our 
tight-binding theory provides a good description for QDQW nanocrystals.

First, we use the empirical tight-binding (ETB) method
\cite{pel89,leung97,leung98,lee01} to perform numerical calculations of 
both electron and hole single-particle energies (i.e, $E_{e}$ and $E_{h}$) 
and eigenstates (i.e., $\mid\Phi_{e}\rangle$ and $\mid\Phi_{h}\rangle$) 
for QDQW's. Our theory can be used to model single or coupled 
nanocrystal systems with spherical, hemi-spherical, tetrahedral 
or pyramidal geometry. In this paper, we focus on single 
spherical QDQW's. Also we assume that atoms in  these nanoparticles 
occupy the sites of a regular fcc lattice. As developed by Vogl, 
Hjalmarson and Dow\cite{vogl83}, each atom has its outer  valence 
$s$ orbital and three outer $p$ orbitals plus a fictitious excited $s^{*}$  
orbital which is included to mimic the effects of higher lying states.
Only on-site and nearest-neighbor couplings between orbitals 
are included in our $sp^{3}s^{*}$ ETB theory. Since the spin-orbital 
coupling is not known for HgS, we shall not consider it  for our 
numerical results reported here. The empirical single-particle 
Hamiltonians are determined by adjusting the matrix elements to 
reproduce known band gaps and effective masses of the bulk band 
structures.  In this paper, our tight-binding parameters for CdS 
and HgS are the same as those of  
Bryant and Jask\'{o}lski\cite{gwb01,bryant02}.
Finally, the electron and hole single-particle eigenstates and energies 
close to the band edges  are  found by diagonalizing the single-particle 
Hamiltonian with an iterative eigenvalue solver. 

\begin{center}
\epsfig{file=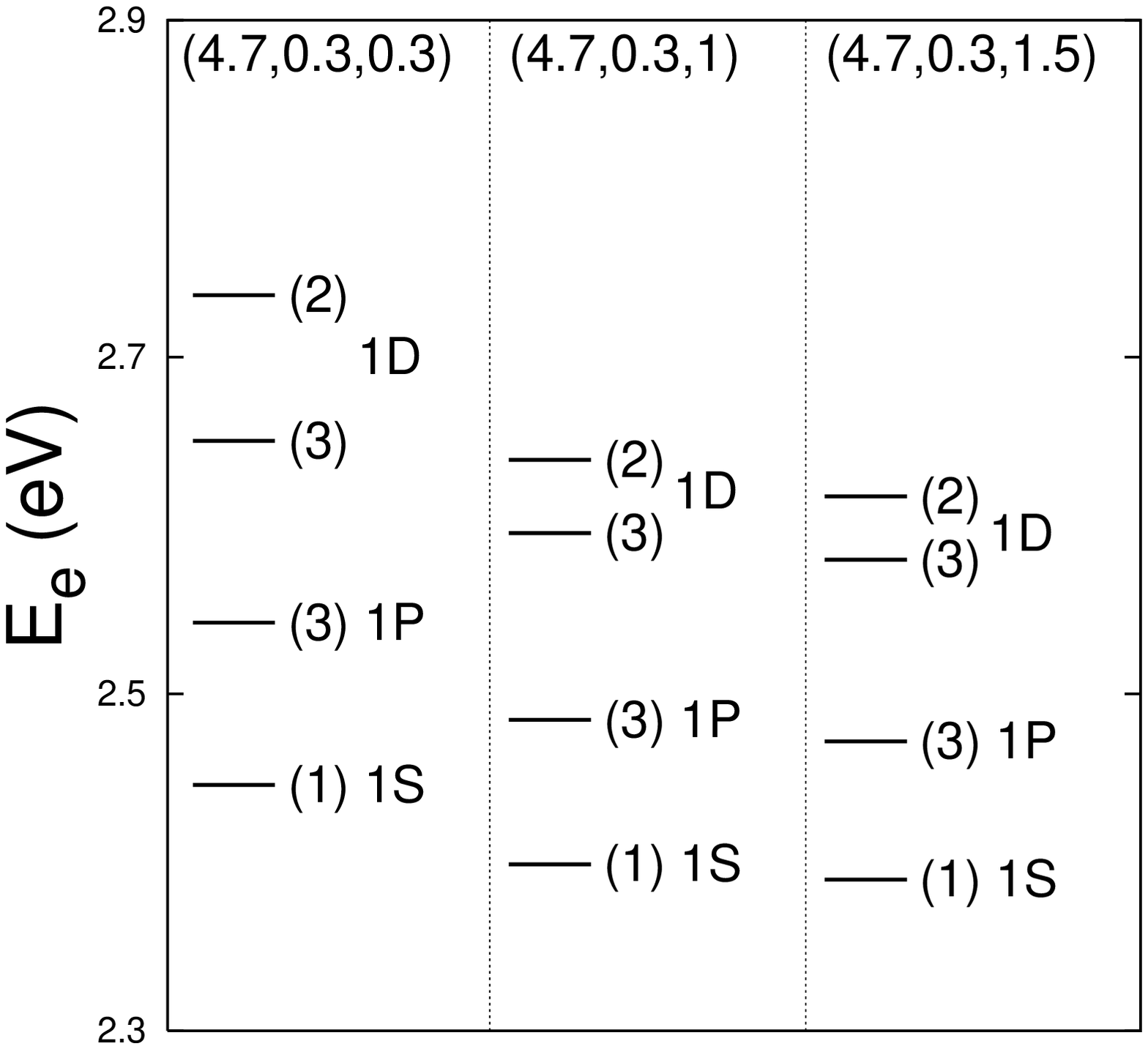,width=8cm,height=6cm}
\epsfig{file=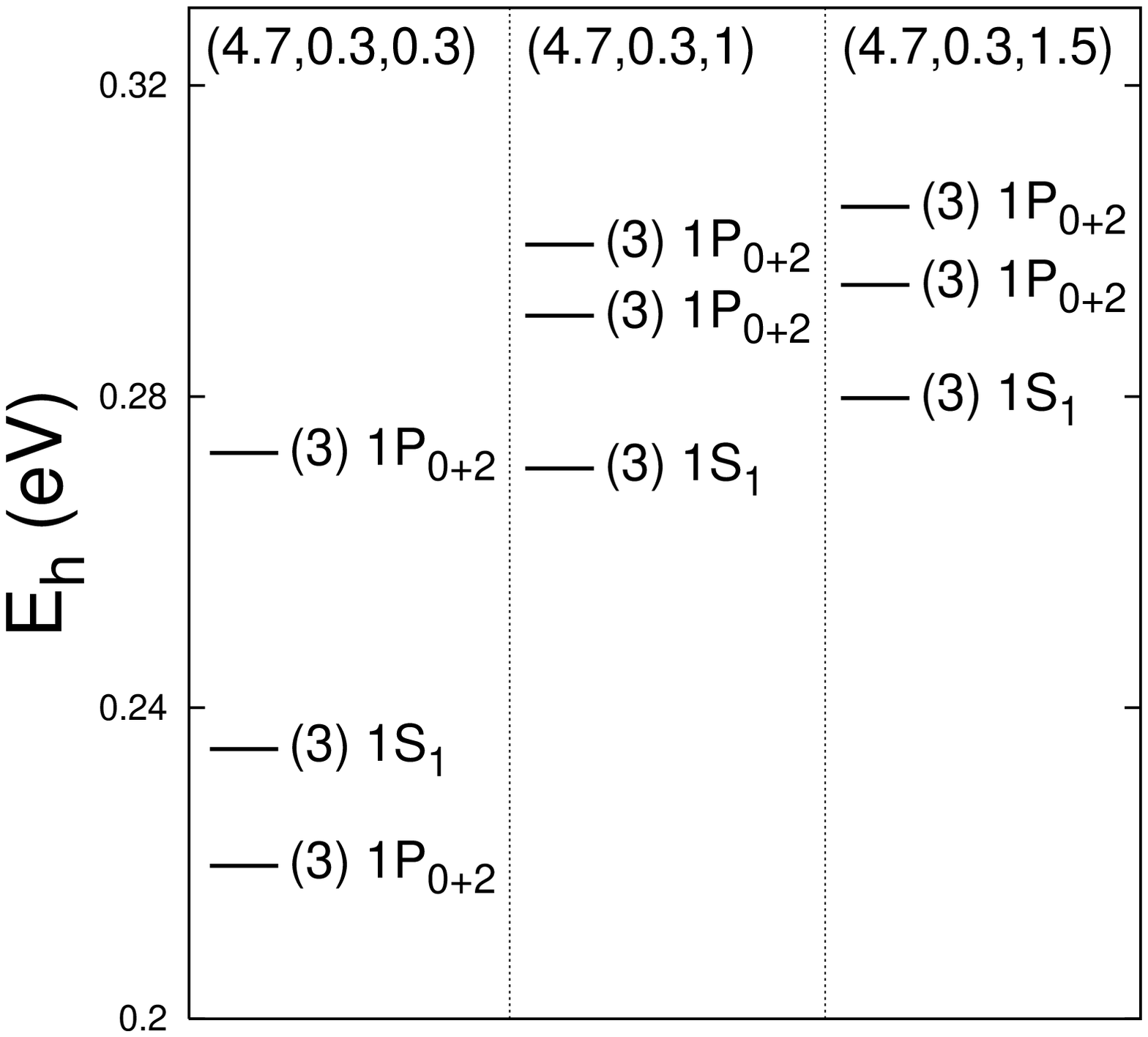,width=8cm,height=6cm}
\end{center}  

\noindent
{\small FIG.1: Electron and hole single-particle energy spectra,
$E_{e}$ and $E_{h}$,  for (4.7, 0.3, 0.3), (4.7, 0.3, 1) and
(4.7, 0.3, 1.5) CdS/HgS/CdS QDQW's. The level degeneracy is shown
in the brackets  and the approximate spherical symmetry of each state
is indicated. Both the electron energy, which increases upward, and the
hole energy, which decreases downward, are referred to the top of
the valence band of CdS.}

\

We describe the effective Hamiltonian of an electron-hole pair by 
combining a two-particle term, $H_{eh}$, which includes  the Coulomb and 
exchange interactions, with a single-particle term, $H_{single}$, which  
contains the kinetic and potential energies of the electron and 
hole\cite{lee01}. The Coulomb and exchange interactions are screened 
by a dielectric function $\epsilon(\mid {\bf r}^{'}-{\bf r}\mid, R)$, 
where $\mid {\bf r}^{'} - {\bf r}\mid$ and $R$ are the separation 
between electron and hole and the quantum dot radius, respectively. 
Our electron-hole basis set $\{\mid\Psi_{i}\rangle\}$ 
is taken by multiplying the spatial part $\mid\Phi_{eh}\rangle$ (namely,  
the product of electron and hole single-particle eigenstates 
$\mid\Phi_{e}\rangle$ and $\mid\Phi_{h}\rangle$)  and their spin states 
$\mid\phi_{spin}\rangle$ (namely,  either the singlet component or one 
of the triplet components of the electron-hole spin state). Then, the 
single-particle  Hamiltonian can be written in terms of the electron-hole 
basis set as 
\begin{equation}
H_{single} = \sum_{i}E_{i}\mid\Psi_{i}\rangle\langle\Psi_{i}\mid,
\end{equation} 
where  $E_{i} = E_{e}^{(i)}-E_{h}^{(i)}$ is the energy difference  
between corresponding electron and hole energies, $E_{e}^{(i)}$ and 
$E_{h}^{(i)}$ of the single-particle Hamiltonian, of the $i$th 
electron-hole pair. Meanwhile,  the electron-hole interaction  
Hamiltonian in the electron-hole basis set is given by 
\begin{equation}
H_{eh} = \sum_{spin}(J+K)\mid\phi_{spin}\rangle\langle\phi_{spin}\mid, 
\end{equation}
where  $J$ and $K$ describe the Coulomb and exchange interactions (the 
same as those of Lee  {\sl et al.}\cite{lee01}). More details about the 
formulas mentioned in this part are given in the work of Lee 
{\sl et al.}\cite{lee01}. 

\

\noindent 
{\small TABLE I: On-site unscreened Coulomb and exchange
integrals, $\omega_{coul}$ and $\omega_{exch}$
for the $sp^{3}s^{*}$ basis set in units of
eV for Cd, Hg and S. Integrals for the $sp^{3}$ orbitals are
calculated based on the hybridized orbitals along the bonding
directions defined by Leung and Whaley \cite{leung97}.}

\begin{center}
\begin{tabular}{ccccc}\hline\hline 
Integral & $(sp_{a}^{3},sp_{a}^{3})$ & $(sp_{a}^{3},sp_{b}^{3})$
&$(sp_{a}^{3},s^{*})$ & $(s^{*},s^{*})$\\ \hline
$\omega_{coul}^{\rm Cd}$  &  5.3346 &  4.0787  & 1.7942 & 1.7181 \\
$\omega_{exch}^{\rm Cd}$  &  5.3346 &  0.5905  & 0.1379 & 1.7181 \\
$\omega_{coul}^{\rm S}$   & 15.5190 & 11.7173  & 3.5295 & 2.8804  \\
$\omega_{exch}^{\rm S}$   & 15.5190 &  1.1836  & 0.0454 & 2.8804  \\
$\omega_{coul}^{\rm Hg}$  &  6.1733 &  4.4216  & 1.9593 & 1.1089 \\
 $\omega_{exch}^{\rm Hg}$ &  6.1732 &  0.6295  & 0.0088 & 1.1089 \\ \hline 
\end{tabular}
\end{center}

\

\noindent 
{\small TABLE II: Coulomb shift $E_{coul}$, exchange splitting
$E_{exch}$  and the first spin-singlet and triplet excitonic energies
  $E_{1}^{(1)}$ and $E_{1}^{(3)}$ for a (4.7, 0.3, 0.3) CdS/HgS/CdS QDQW.
$N_{e}$ and $N_{h}$ denote the number of electron and hole single-particle
states included in the basis set.}
\begin{center}
\begin{tabular}{ccccrc}\hline\hline
$N_{e}$            &
$N_{h}$            &
$E_{1}^{(1)}$      &
$E_{1}^{(3)}$      &
$E_{coul}$         &
$E_{exch}$         \\
    &     & [ meV ]   & [ meV ] & [ meV ] & [ meV ]  \\ \hline
  1 &  3  &  2079.62  &2079.11  & 94.01   & 0.51  \\
  4 &  3  &  2073.23  &2072.46  & 100.66  & 0.77  \\
  4 &  6  &  2067.28  &2066.22  & 106.90  & 1.06  \\
  4 &  9  &  2067.19  &2066.12  & 107.00  & 1.07 \\ \hline
\end{tabular}
\end{center}

\

As introduced by Leung and Whaley\cite{leung97} and 
Lee {\sl et al.}\cite{lee01}, the Coulomb and exchange interaction 
matrix elements are expressed in terms of the Coulomb and exchange 
integrals, $\omega_{coul}$ and $\omega_{exch}$, of our ETB orbitals. 
Table I lists the unscreened on-site Coulomb and exchange integrals 
for the $sp^{3}s^{*}$ basis set for  Cd, S and Hg  calculated by 
 using a Monte Carlo method with importance sampling for the radial 
integrations\cite{lee01}. Regarding off-site Coulomb integrals, we 
 estimate them using the  Ohno formula modified by Leung and Whaley
\cite{leung97}.  It is known that off-site exchange integrals decrease 
quickly as the distance between atom sites increases due to the 
localization and orthogonality of orbitals\cite{lee01}. Here, we use 
the off-site exchange integrals of Leung, Pokrant and 
Whaley\cite{leung98}.

The high-frequency dielectric constant \cite{ds94} for the different 
regions of CdS/HgS/CdS QDQW's are $\epsilon_{CdS}=5.5$, 
$\epsilon_{HgS}=11.36$, and $\epsilon_{H_{2}O}=1.78$. In our 
numerical calculations, we use an average, effective high-frequency 
dielectric constant, $\epsilon_{ave}=6$.

\

\begin{center}
\epsfig{file=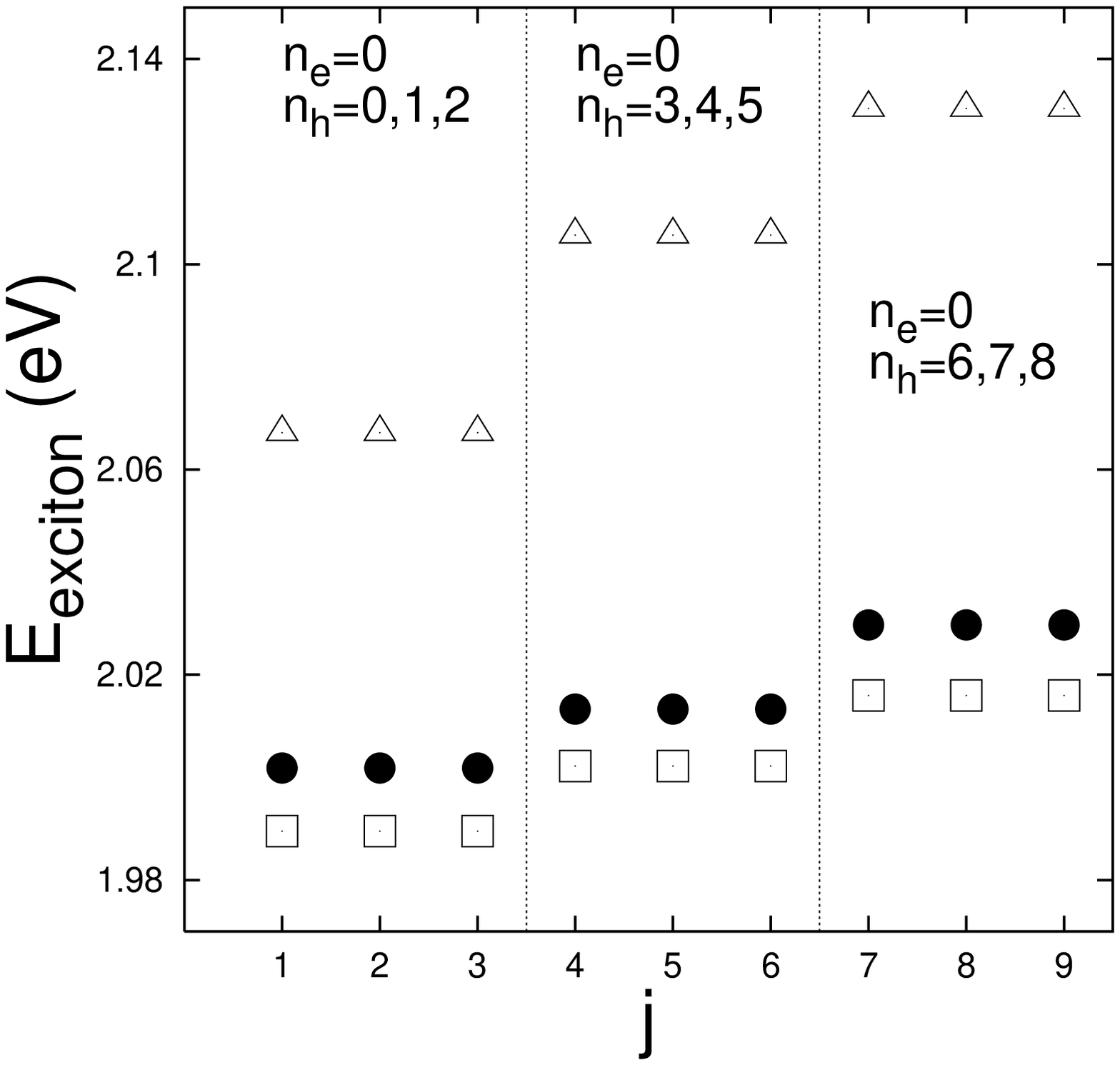,width=8cm,height=6cm}
\end{center}  

\noindent
{\small FIG.2: Spin-singlet excitonic energy spectrum for
the first few excitonic states ($j$,  the state
serial number), where $\triangle$, $\bullet$ and
$\Box$ are for (4.7, 0.3, 0.3), (4.7, 0.3, 1) and
(4.7, 0.3, 1.5) CdS/HgS/CdS QDQW's, respectively. The
primary electron-hole pair states making up each of
the triply degenerate excitonic levels are indicated.}

\

\begin{center}
\epsfig{file=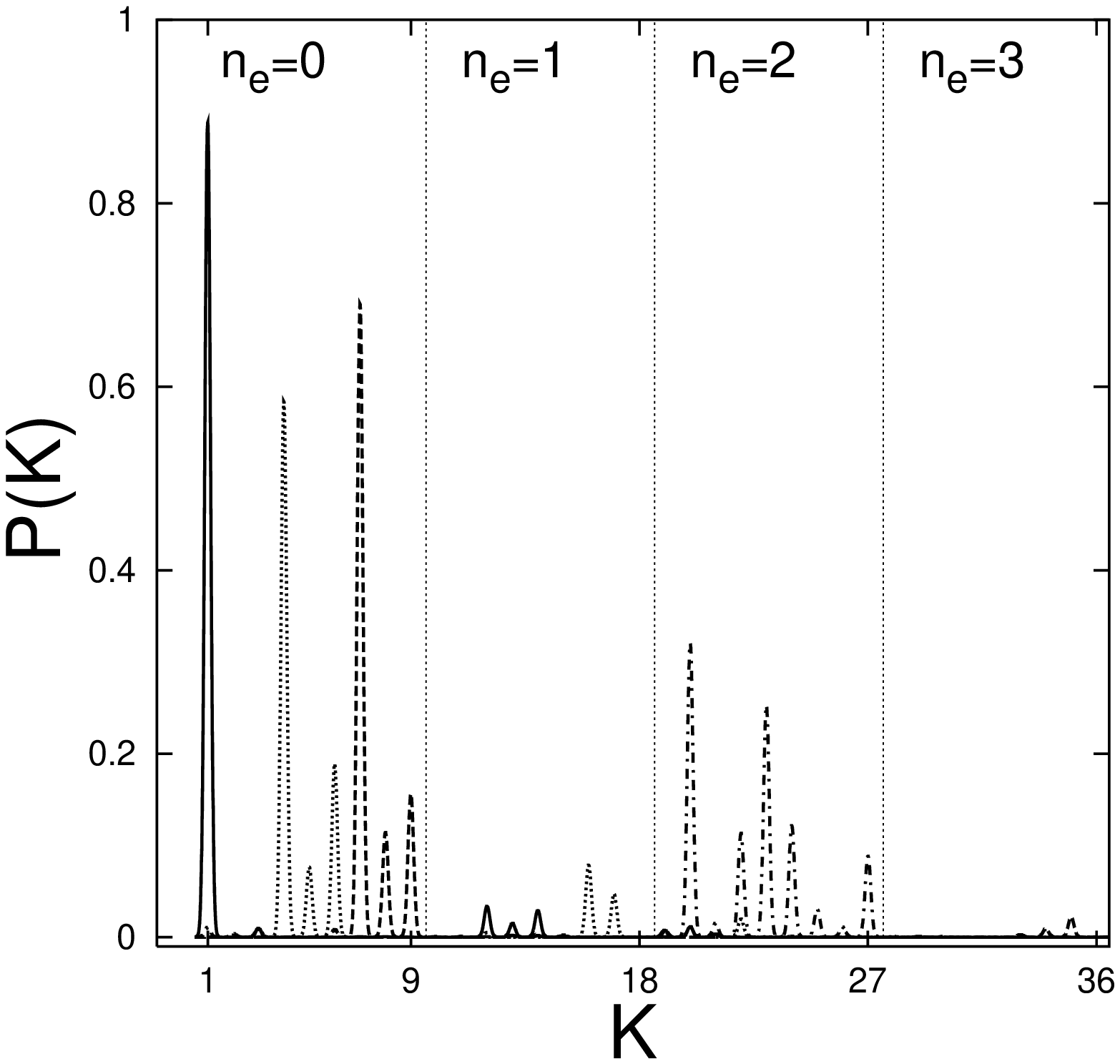,width=8cm,height=6cm}
\end{center}
 
\noindent
{\small FIG.3: Occupation probability $P(K)$ of the
$K$th electron-hole pair $(n_{e},n_{h})$  in the
first (solid line), 5th (dotted line),
8th (dashed line)  and 22nd (dot-dashed line)
spin-singlet excitonic states for the (4.7, 0.3, 0.3)
CdS/HgS/CdS QDQW. $P(K)$ is broadened by  a
Gaussian to enhance visualization.}

\

Because of the total spin of the electron-hole pair, we have two 
Hamiltonians: one for a spin singlet including both the Coulomb 
and exchange interactions, and another for a spin triplet having  
only the Coulomb interaction. Therefore, we diagonalize both 
Hamiltonians in the electron-hole basis separately and obtain a 
set of spin-singlet and triplet excitonic states. For clarity, we 
use the parameter set ($D_{core}$, $L_{well}$, $L_{clad}$) 
to denote a CdS/HgS/CdS QDQW, where $D_{core}$, $L_{well}$ and 
$L_{clad}$  are the CdS core diameter, HgS well width and CdS 
clad thickness, respectively, in nm. Also we mention  three 
definitions of Lee {\sl et al.}\cite{lee01}: 
(i) {\sl optical excitonic gap} $E_{g}^{opt}$ is the lowest 
spin-singlet excitonic energy $E_{1}^{(1)}$; (ii) the difference 
between the single-particle energy gap $E_{g}^{single}$ 
and  the lowest spin-triplet excitonic  energy $E_{1}^{(3)}$ 
is defined as the {\sl Coulomb shift}, namely, 
$E_{coul}=E_{g}^{single}-E_{1}^{(3)}$; (iii)  the difference 
between the lowest spin-singlet and triplet excitonic energies 
is defined as the {\sl exchange splitting}, namely, 
$E_{exch}=E_{1}^{(1)}-E_{1}^{(3)}$. 
 
For the band gaps of HgS and CdS, we use 
${\rm E}_{\rm g, HgS} = 0.2$ eV and 
${\rm E}_{\rm g, CdS} = 2.5$ eV, respectively, with the CdS conduction 
band edge 1.45 eV above the conduction-band edge of HgS\cite{wj98}. 
Fig.1  shows the electron and hole single-particle energy spectra of 
(4.7, 0.3, 0.3), (4.7, 0.3, 1) and (4.7, 0.3, 1.5) CdS/HgS/CdS QDQW's, 
which have been characterized experimentally by 
Weller and coworkers\cite{ds94}. The calculated single-particle 
energy gaps are 2.173 eV, 2.099 eV and 2.085 eV, respectively. 
The degeneracy is also shown in the brackets. The lowest electron 
and hole states can be described approximately by the 
spherical symmetries indicated in 
Fig.1, i.e., 1S-, 1P- and 1D-like electron states, mixed 
${\rm 1P_{0} + 1P_{2}}$-like hole states, and a ${\rm 1S_{1}}$-like 
hole state. Here, S, P, D indicate the spatial angular momentum and
the subscript for holes indicates the total angular momentum sum of
the spatial and atomic orbital angular momenta. As the clad thickness 
increases, the level ordering of the electron states does not change, 
but that of the last two hole states switches (as shown in Fig.1). 
The ordering of hole levels switches because excited P hole levels 
are the states most sensitive to the potential far from the dot center.

In this work, we focus mainly on the lowest singlet and 
triplet excitonic energies. In our electron-hole basis set, 
we include the 4 electron ($n_{e} = 0, 1, 2, 3$) and 9 hole 
($n_{h}=0, 1 , ... , 8$) single-particle states closest to 
the band edge, where $n_{e}$ (labelling initially from the 
ground electron state) and $n_{h}$ (labelling initially 
from one of the triply degenerate ground hole states) 
are the indices for electron and hole single-particle states. 
For example, Table II lists  the Coulomb shift, exchange 
splitting and the lowest  spin-singlet and triplet  
excitonic energies for the (4.7, 0.3, 0.3) QDQW by considering 
different numbers of electron and hole single-particle states 
close to the band edge. The cases shown are chosen to 
ensure that all states of a given degenerate level are 
included. The inclusion of more electron-hole 
configurations leads to an increase of the Coulomb  shift and 
exchange splitting. The excitonic energies are converged 
within a few meV when $N_{e} = 4$ and $N_{h} = 9$. 

Fig.2 shows the singlet excitonic energy spectrum for the lowest
few excitonic states in the three QDQW's. For the electron-hole
basis,  $N_{e}=4$ and $N_{h}=9$, the first three triply
degenerate excitonic energy levels come mainly from the contribution 
of electron-hole pairs formed with the ground-state electron, as 
indicated in Fig.2. Other higher excitonic energy levels are due 
to electron-hole pairs made from the excited-state electron. 
This is clearly shown in Fig.3, which presents the occupation 
probability $P(K)$ of the $K$th electron-hole pair $(n_{e}, n_{h})$,
 $K \equiv (n_{e}, n_{h}) = n_{h} +1 + 9 n_{e}$, for one exciton state 
from each of the three lowest triply degenerate exciton levels (the 
first, 5th and 8th exciton states in Fig.2) and a higher exciton 
level (the 22nd exciton state). Fig.3 shows that correlation does not 
strongly mix pair states of different energy in forming the low 
excitonic states. This explains why convergence is achieved with 
only a few basis states. Our previous calculations of correlation 
effects in QDQWs\cite{gwb95} showed that correlation effects are 
weaker in QDQWs than in quantum dots. The extra local confinement 
of the electron and hole to the quantum well inside the quantum dot
increases the splitting between the lowest single-particle states 
and suppresses Coulomb mixing of these states.

\

\noindent 
{\small TABLE III: The Coulomb shift $E_{coul}$, exchange splitting $E_{exch}$,
optical excitonic gap $E_{g}^{opt}$, and 1S-1S transition energy
$E_{1S-1S}$  calculated for (4.7, 0.3,
$L_{clad}=0.3, 1, 1.5$ ) CdS/HgS/CdS QDQW's. $E_{g}^{exp}$ is
the band gap measured by Weller's group\cite{ds94}.
$E_{coul}^{weller}$ and $E_{\rm 1S-1S}^{weller}$ are
the Coulomb shift and  ${\rm 1S-1S}$ transition energy,
respectively, calculated with the effective mass approximation.}
\begin{center}
\begin{tabular}{cccccccc}\hline\hline
${\small L_{clad}}$              &
${\small E_{coul}}$              &
${\small E_{exch}}$              &
${\small E_{1S-1S}}$             &
${\small E_{g}^{opt}}$           &
${\small E_{g}^{exp}}$           &
${\small E_{1S-1S}^{weller}}$    &
${\small E_{coul}^{weller}}$   \\
 $[ {\rm nm} ]$ & [ meV ]
& [ meV ]  & [eV] & [ eV ]  & [ eV ] & [ eV ]  & [ meV ] \\ \hline
0.3    & 107.00  & 1.07 &2.10566   &2.06719  & 2.10   & 2.27    & 84 \\
1.0    & 98.33   & 0.88 &2.01329   &2.00183  & 1.96   & 2.16    & 76\\
1.5    & 96.66   & 0.85 &2.01595   &1.98951  & 1.94   & 2.15    & 75 \\ \hline
\end{tabular}
\end{center}
 
\

Table III summarizes  our calculated  Coulomb shift
$E_{coul}$, exchange splitting $E_{exch}$, 1S-1S transition 
energy $E_{1S-1S}$ and  optical excitonic gap $E_{g}^{opt}$ 
for the three CdS/HgS/CdS QDQW's.  
$E_{1S-1S}$ is the energy of the lowest optically active 
transition because it is a transition between the 
even parity 1S electron state and the odd parity $1{\rm S}_{1}$ 
hole state. $E_{g}^{opt}$ is the energy of lowest possible 
spin-singlet transition which is between the lowest electron 
(even parity 1S) and lowest hole (even parity $1{\rm P}_{0}$). 
$E_{g}^{opt}$ corresponds to the emission peak in experiment.  
In the experiment, the points of maximum curvature in the 
absorption spectra are a good measure 
for the 1S-1S transition energies\cite{weller90} and the 
experimentally measured gaps $E_{g}^{exp}$\cite{ds94} for the 
three CdS/HgS/CdS QDQW's are listed in Table III. We find that our 
calculations for 1S-1S exciton energies are in good agreement with 
experimental $E_{g}^{exp}$ (the relative errors between the experiment 
and our theory are $0.2\ \%$, $2.7\ \%$ and $3.9\  \%$ for 
$L_{clad}=$ 0.3 nm, 1 nm, 1.5 nm, respectively). Also Table III shows that
our calculated Coulomb shift varies from 107 meV  to 96 meV and
the exchange splitting from 1 meV to 0.8 meV when $L_{clad}$ varies
from 0.3 nm to 1.5 nm. 

Including a finite barrier to represent the water solution
 and the Coulomb interaction between electron and hole,
 Weller and coworkers\cite{ds94} used an effective mass model 
to calculate the $1S-1S$ transition energy $E_{1S-1S}^{weller}$ 
and Coulomb shifts $E_{coul}^{weller}$ for the three 
CdS/HgS/CdS QDQW's. Their findings are also listed in 
Table III.  The relative errors between their calculated 1S-1S 
transition energies and experimentally measured ones are
8 $\%$, 10 $\%$ and 11 $\%$ for $L_{clad}=$ 0.3 nm, 1 nm,
1.5 nm, respectively. The ETB theory clearly provides
more accurate energy  gaps than the effective  mass model.
The predicted gaps of the ETB differ slightly from the
measured gaps but are about 150 meV  less than the prediction of
the effective mass model. Most of these differences are due to
differences in the single particle energies. The remaining part 
of the difference in energy gaps is due to the difference in 
Coulomb shifts. The Coulomb shifts predicted by effective mass 
theory are 20 meV lower than the ETB results. It should also 
be noted that single-band effective theory\cite{ds94,gwb95} 
predicts the optically active 1S-1S transition to be the ground 
state transition. Experiment\cite{am96} and tight-binding theory
\cite{gwb01,bryant02}  show that the ground state is dark and 
must be the 1S-1P transition as we have calculated for the exciton 
ground state.

In the tight-bind model, the spin-orbit interaction is determined 
by the parameter $\lambda_{i}=\langle x_{i}, \uparrow\mid H_{so} 
\mid z_{i}, \downarrow\rangle$ where $i=a$ (anion) or 
$c$ (cation) and $H_{so}$ is the Hamiltonian of spin-orbit 
interaction\cite{chadi77,hass83}. In the bulk, it is known that 
the spin-orbit coupling lifts the degeneracy of the zone-center 
bulk band states and produces a 4-fold degenerate state 
(the light and heavy hole bands) 
and a 2-fold degenerate state (the split-off band) at the zone center. 
For example, in bulk CdS, the zone-center splitting between the 
split-off band and the light and heavy hole bands is about 80 meV\cite{wj98}. 
 Bryant and Jask\'{o}lski\cite{bryant02} have shown that the spin-orbit 
splittings of single-particle states in the QDQWs are much smaller than 
the bulk zone-center ones. 
At large wavevector, the bulk CdS and HgS band  structures show little 
effect of spin-orbit coupling. Spin-orbit effects and mixing are 
weak in QDQWs because the strongly confined trap 
states are made from bulk states with large wavevector. 
Coulomb interaction doesn't strongly mix different levels, so 
our results should not change much when we include spin-orbit interaction. 
Detailed work about the effect of spin-orbit interaction 
on the fine structure of the excitonic spectrum is still in 
progress and will be reported in a future paper. 
 
The ETB model we have considered for the QDQW has the following 
specific features: 
a spherical geometry, a particular well thickness and position 
inside the QDQW, no faceting of the surfaces or interfaces, 
a perfect fcc lattice, specific choices for uncertain 
material parameters (e.g., valence band offset, dielectric 
constant $\epsilon$ ), and water barrier not included. 
However, the electronic structure obtained 
for the QDQW is determined mainly by the trapping and state 
symmetry and does not appear to depend significantly on a 
precise choices made specifically for QDQW geometry and material 
parameters. Bryant and 
Jask\'{o}lski\cite{bryant02} have test this by studying single-particle 
states in QDQWs with different shapes and different well geometries  
and they have determined the dependence of the results on valence band offsets 
and spin-orbit coupling. For example, they have found that 
similar electron (hole) states exist for spherical, tetrahedral 
and cutoff tetrahedral QDWSs provided that a similar number of 
cations (anions) occupy a particular region for each shape; 
the splitting between electron (hole) levels has a weak 
dependence on well thickness, well position, valence 
band offset, or as already mentioned, spin-orbit effects. 
The main effect of any of these uncertainties would be on 
the absolute energy position of the ground state transition. 
The position of other transitions, relative to the ground state 
transition, are insensitive to the uncertainties. Our good 
agreement with experiment shows that we also describe the 
lowest transition well with our model. The computational tests 
of Bryant and Jask\'{o}lski\cite{bryant02} are not 
exhaustive.  Other possibilities (e.g., faceting at surfaces or 
interfaces, dependence on other tight-binding parameters, the 
dielectric screening, or on the effect of the water barrier) 
could be considered. However, further experimental characterization of 
QDQW geometry and material parameters and reduced experimental 
uncertainty are needed to provide tighter
constraints for more complete sensitivity tests. 
 
In summary, we have studied electron-hole correlations  in 
QDQWs by incorporating Coulomb and exchange interactions into 
an empirical tight-binding model. Our calculated 
optical excitonic gaps are in good agreement with the 
experiment. Our ETB theory can provide a good description for 
QDQW's nanocrystals. 



\begin{references}

\bibitem{ph00}P. Harrison, {\sl Quantum  Wells, Quantum Wires and
 Quantum Dots} (John Wiley \& Sons, New York, 2000).

\bibitem{kbar01}K. Barnham and D. Vvedensky, {\sl Low-dimensional 
Semiconductor Structures: Fundamentals and Device Applications} 
(Cambridge University Press, New York, 2001).

\bibitem{ds94}D.Schooss, A. Mews, A. Eychm\"{u}ller, and H. Weller, 
Phys. Rev. B {\bf 49}, 17072 (1994).

\bibitem{am94}A. Mews, A. Eychm\"{u}ller, M. Giersig, D. Schooss, and 
H. Weller, J. Phys. Chem. {\bf 98}, 934 (1994).

\bibitem{am96}A. Mews, A. V. Kadavanich, U. Banin, and A. P. 
Alivisatos, Phys. Rev. B {\bf 53}, R13242 (1996).


\bibitem{svk99}S. V. Kershaw,  M. Burt, M. Harrison, A. Rogach, 
H. Weller, and A. Eychm\"{u}ller, Appl. Phys. Lett. {\bf 75}, 1694 (1999).

\bibitem{rbl2001}R. B. Little, M. A. El-Sayed, G. W. Bryant, 
and S. Burke, J. Chem. Phys. {\bf 114}, 1813 (2001).

\bibitem{jwh93}J. W. Haus, H. S. Zhou, I. Honma, and H. Komiyama, 
Phys. Rev. B {\bf 47}, 1359 (1993).

\bibitem{gwb95}G. W. Bryant, Phys. Rev. B {\bf 52}, R16997 (1995).

\bibitem{wj98}W. Jask\'{o}lski and  G. W. Bryant, Phys. Rev. B {\bf 57}, 
R4237 (1998).

\bibitem{gwb01}G. W. Bryant and W. Jask\'{o}lski, Physica E {\bf 11}, 72 (2001).
 
\bibitem{bryant02} G. W. Bryant and W. Jask\'{o}lski, ``Tight-binding
theory of quantum-dot quantum wells: single particle effects and near band-edge
structure", to be submitted to Phys. Rev. B.

\bibitem{pel89}P. E. Lippens and M. Lannoo, Phys. Rev. B {\bf 39}, 10935 (1989).

\bibitem{leung97} K. Leung and K. B. Whaley, Phys. Rev. B {\bf 56}, 7455 (1997).

\bibitem{leung98} K. Leung, S. Pokrant, and K. B. Whaley, Phys. Rev. B 
{\bf 57}, 12291 (1998).

\bibitem{lee01}S.  Lee, L. J\"{o}nsson, J. W. Wilkins,  G. W. Bryant, 
and G. Klimeck, Phys. Rev. B {\bf 63}, 19518 (2001).

\bibitem{vogl83} P. Vogl, H. P. Hjalmarson, and J. D. Dow, J. Phys. Chem. 
Solids {\bf 44}, 365 (1983).

\bibitem{chadi77} D. J. Chadi, Phys. Rev. B {\bf 16}, 790 (1977).

\bibitem{hass83} K. C. Haas, H. Ehrenreich, B. Velick\'{y},  
Phys. Rev. B {\bf 27}, 1088 (1983).

\bibitem{weller90}L. Katsikas, A. Eychm\"{u}ller, M. Giersig, and
H. Weller, Chem. Phys. Lett. {\bf 172}, 201 (1990).


\end{references}
\end{document}